\documentstyle[12pt,aasms4]{article}

\begin{document}

\title{ 
K-Band Spectra and Narrowband Photometry of DENIS Field 
Brown Dwarfs
}

\author{A.T. Tokunaga\altaffilmark{1,2}
\affil{Institute for Astronomy, University of Hawaii,\\
2680 Woodlawn Drive, Honolulu, HI  96822}
\author{N. Kobayashi\altaffilmark{1,2,3}}
\affil{Subaru Telescope, National Astronomical Observatory of Japan,\\
650 North A$'$ohoku Place, Hilo, HI  96720}  }

\altaffiltext{1}{Visiting Astronomer, United Kingdom Infrared Telescope.
The United Kingdom Infrared Telescope is operated by the Joint Astronomy
Centre on behalf of the U.K. Particle Physics and Astronomy Research 
Council.} 
\altaffiltext{2}{Visiting Astronomer at the Infrared Telescope Facility,
which is operated by the University of Hawaii under contract from the 
National Aeronautics and Space Administration.} 
\altaffiltext{3}{Visiting Astronomer, Institute for Astronomy, University
of Hawaii.}

\begin{abstract}
Infrared spectra at 1.9-2.5 $\mu$m and narrowband photometry of three
low-mass objects,  DENIS-P J0205.4--1159, J1058.7--1547, and 
J1228.2--1547, are presented.
As shown previously by Delfosse et al. (1997, A\&A, 327, L25),
DENIS-P J0205.4--1159 shows an absorption feature at 2.2 $\mu$m.
We attribute this absorption to H$_2$.
A simple two-parameter analysis of the K-band spectrum of low-mass objects 
is presented in which the relative strength of the H$_2$O and H$_2$ 
absorption bands is found to be correlated with the effective temperature 
of the objects.
The analysis confirms that DENIS-P J0205.4--1159 is the lowest-temperature 
object of the three.
We present narrow-band photometry of these objects which provides the
continuum flux level inbetween the deep H$_2$O absorption bands.
These data show the continuum level accurately for the first time,
and they will provide tight constraints for spectral models of these 
interesting objects.

\end{abstract}


\section{Introduction}

The discovery of brown dwarfs in the field by infrared surveys as well as 
by a proper motion survey has demonstrated both the existence and the 
abundance of such objects.  Three brown dwarf candidates, DENIS-P 
J0205.4--1159, J1058.7--1547, and J1228.2--1547 were reported by 
Delfosse et al. (1997).  
(For simplicity we refer to these objects as D02--11, D10--15, and 
D12--15.)  
The detection of lithium in D12--15 suggests that it is a brown dwarf
(Mart\'{\i}n et al. 1997; Tinney, Delfosse, \& Forveille 1997);
however, D10--15 did not show any lithium in its spectrum.  
There were no observations searching for lithium in D02--11 reported at the 
time this paper was written,
although the spectrum of D02--11 shows that it is the coolest object of the 
three and therefore it most likely is a brown dwarf (Delfosse et al. 1997; 
Tinney et al. 1997).
Another  field brown dwarf, Kelu-1, was discovered in a proper motion 
survey (Ruiz, Leggett, \& Allard 1997).
The presence of lithium in this object, its estimated effective temperature, 
and its absolute magnitude  all indicate that it is a brown dwarf.

In this paper, we report on new $K$-band spectra of D02--11, D10--15, and 
D12--15, as well as narrowband photometry at 1.28--3.8 $\mu$m.

\section{Observations}

Spectra at 1.9--2.5 $\mu$m were obtained with the CGS-4 infrared 
spectrograph at the UKIRT on 1997 December 23--24.  
The 40 lines mm$^{-1}$ grating used provides a spectral resolution of 900 
at 2.2 $\mu$m with a 0.6$''$ slit.
For the removal of the telluric absorption lines, we observed the bright 
stars HR 692 (F0V), HR 831 (F6III--IV), and HR 1978 (F0III).
The spectra were obtained by oversampling by a factor of 3.
This was achieved by moving the detector array one-third of a resolution 
element between exposures. 
During the observations every fourth column of the array was unusable
due to an electronic problem.
We were able to recover the spectrum because each resolution element
was sampled with three grating positions.
However a ripple that was caused by the rejection of every fourth column
remained in the spectrum.
This ripple was eliminated by applying a Fourier Transform to the spectrum, 
removing the frequencies corresponding to the ripple, 
and then applying the inverse Fourier Transform.
The ripple was not evident after this processing, and the division by a 
standard star spectrum showed no artifacts from the bad columns of the 
array.
At the final stage of reduction, the spectra were smoothed with a 3-pixel 
boxcar function that matched the spectral resolution of the spectrograph.
The spectra are shown in Figure 1.
\placefigure{fig1} 

Narrowband filter photometry at 1.28, 1.68, and 2.18 $\mu$m and at 
L$'$ (3.76 $\mu$m) were obtained with the NSFCam infrared camera at the 
IRTF on 1997 December 28--29 UT.
The observations at 1.28 $\mu$m were obtained with a narrowband 1\% 
filter and at 1.68 and 2.18~$\mu$m using a 1\% circular variable filter.
Photometric standard stars HD106965 and HD18881 were used 
(Elias et al. 1982).
The results are summarized in Table 1.
The uncertainty in the observations is about 2\%.
\placetable{table1}

For D10--15 and D12--15 the flux density was consistent with that determined 
from the spectral standard (HR 1978) and by using a $V-K$ color 
appropriate for the spectral type of the standard.
For D02-11 there was a 22\% discrepancy between the narrowband photometry 
and the spectral standard star.  
This discrepancy may have arisen from the $K$ magnitude 
uncertainty of the spectral standard stars (HR 692 and HR 831) or
from loss of light at the slit.
Therefore we renormalized the spectrum to the flux density obtained 
from the narrowband photometry.

\section{Discussion}

\subsection{Spectroscopy}

Spectra presented by Mart\'{\i}n et al. (1997) and Tinney et al. (1997) 
suggest that the order of warmer to cooler is: 
D10--15, D12--15, and D02--11.
The spectra are shown in this order in Figure 1.
Strong H$_2$O absorption at 2.0 $\mu$m and CO absorption at 2.3 
$\mu$m are evident.
The overall spectral shapes of D10--15 and D12--15 are similar but that of 
D02--11 is quite different in the following aspects:
(1) There is a steeper rise of the spectrum from 2.0--2.1 $\mu$m, and
(2) There is a ``dip'' at 2.1--2.3 $\mu$m compared with D10--15 and 
D12--15 (more easily seen in Fig. 2).
The dip is also seen in the spectrum of D02--11 presented by 
Delfosse et al. (1997).

Closer inspection of the spectra shows additional spectral differences:
(1) There is an absorption feature at 2.01 $\mu$m in D02--11 that is not 
seen in D10--15 and D12--15.
(2) There is a difference in the slope of the spectrum at 1.95--2.0 $\mu$m.
D02--11 is flat, but D10--15 and D12--15 are both rising toward shorter 
wavelengths.
This appears to be a result of deeper H$_2$O absorption in D02--11.
(3) There is a marginally significant absorption feature at 
2.200$\pm$0.002 $\mu$m in the spectrum of D02--11 that is slightly 
offset from the NaI doublet at 2.206 and 2.209 $\mu$m. 
This absorption feature was also observed by Delfosse et al. (1997). 
A detailed spectral model is needed to confirm the identity of this
absorption feature.
There are no atomic lines listed by Jones et al. (1994) that are 
clearly seen in our spectra.
(4)~There are perhaps many weak absorption features throughout the 
spectrum.  The reality of these features requires both modeling and 
additional spectra for confirmation.  We cannot rely on standard 
signal-to-noise arguments in assessing the reality of these weak features
because of the detector array problem mentioned in \S 2.

\subsubsection{ Comparison to Other Low-Temperature Objects}

Figure 2 shows the comparison of D02--11 to other low-temperature 
objects Kelu-1 and GD165B. 
\placefigure{fig2}
The spectra were arbitrarily normalized at 2.07 $\mu$m.
This wavelength was chosen as it is intermediate between the maximum of the
H$_2$O and the H$_2$ absorption bands.
Spectra of the latter two objects were taken from Ruiz et al. (1997) and 
Jones et al. (1994).
Kelu-1 is another recently discovered field brown dwarf with detected Li 
absorption (Ruiz et al. 1997).  
GD165B is an unusual low-temperature object and is a likely brown dwarf 
(Kirkpatrick, Henry, \& Liebert 1993, Jones et al. 1994). 
Unfortunately it is not possible to observe the Li absorption line at 670.8 
nm in this object because of strong scattered light from the primary star.
Note the strong similarity of the spectrum of both Kelu-1 and GD165B.
As in the comparison to D10--15 and D12--15, the spectra of Kelu-1 and 
GD165B are different from that of D02--11 by having relatively higher flux 
density at 1.9--2.0 $\mu$m and at 2.15--2.30 $\mu$m.

Figure 3 shows the comparison of D02--11 to D10--15 and GL229B.  
\placefigure{fig3}
The spectra shown in Figure 3 seem to imply a progression of deeper 
H$_2$ and CH$_4$ absorption, with the spectrum of D02--11 being 
intermediate between that of D10--15 and GL229B.
This suggests that the effective temperature of 
D02--11 is less than D10--15, a result consistent with that of
Tinney et al. (1997).
We note that the 1.9--2.5 $\mu$m spectrum of D10--15 is very similar to 
that of VB 10, an M8 V spectral type star, at 1.9--2.5 $\mu$m.
However these objects are known to have significant spectral differences at 
0.65--0.90 $\mu$m (Tinny et al. 1997).
Therefore the 1.9--2.5 $\mu$m spectrum alone is not sufficient to 
precisely classify objects.  
Nonetheless, we show in the next section that the 1.9--2.5~$\mu$m 
spectrum may be sufficient to enable a rough classification of objects and 
to identify brown dwarfs.

Compared to the other objects, the spectrum D02--11 exhibits a broad
and  shallow dip at 2.2~$\mu$m (see the comparison in Figure 3).
We believe this broad absorption is caused by H$_2$ absorption based
on a comparison of our spectra to atmospheric models of low-mass objects 
calculated by T. Tsuji (1998).
These models show that H$_2$ is the major absorber in this spectral region 
and that CH$_4$ is a minor absorber for an effective temperature of about 
1800 K. 
This is consistent with Ruiz et al. (1997), 
who indicate that H$_2$ is the main cause of the absorption
at 2.2~$\mu$m in the brown dwarf Kelu--1 (see their Fig. 3).
Additional modeling is needed to determine the best fitting parameters for 
D02--11, but this is beyond the scope of this paper.
Note that Delfosse et al. (1997) attribute the dip at 2.2 $\mu$m to CH$_4$, 
and as a result they estimate the effective temperature to be about 1500~K.
However no support for this interpretation was presented,
and we assume our effective temperature estimate in this paper.

\subsubsection{ Classification by Means of a``Slope'' Analysis}

The comparison of D02--11 to various objects leads to the question of 
whether a crude type of classification of objects might be possible using 
the shape of the spectrum.
We define two parameters as follows:

\begin{equation}
K1 = \frac{  <F_{2.10-2.18}> - <F_{1.96-2.04}>  } 
    {  0.5 \{ <F_{2.10-2.18}> + <F_{1.96-2.04}> \}  }
\end{equation}

\begin{equation}
K2 = \frac{  <F_{2.20-2.28}> - <F_{2.10-2.18}>  } 
    {  0.5 \{ <F_{2.20-2.28}> + <F_{2.10-2.18}> \}  }
\end{equation}

\noindent where $< F_{w1-w2} >$ is the average flux density in the 
wavelength 
range $w1$ to $w2$.  
The label  ``$K$'' is used because the spectra used are in the 
photometric $K$-band.

These definitions are similar to those of Steele et al. (1998) for the 2.1 
$\mu$m slope with the exception that we base our definition based on 
F$_\lambda$ versus wavelength in microns instead of F$_\nu$ versus 
wavelength.
We also define the denominator to be the average instead of the sum.

$K1$ measures the rise of the spectrum from 2.0 to 2.14 $\mu$m, 
and this is primarily affected by the amount of H$_2$O absorption.
$K2$ measures the amount of absorption between 2.14 and 2.24 $\mu$m 
that is caused by H$_2$.
For extremely low-effective temperature objects,
$K2$ will be sensitive to the amount of CH$_4$ as well.
In D02--11, D10--15, and D12--15 the K2 index is influenced primarily 
by the amount of H$_2$ absorption while in an extremely cool object 
such as GL229B it is affected primarily by CH$_4$.
This results from the overlapping absorption of H$_2$ and CH$_4$ absorption 
bands at about 2.2 $\mu$m.
As a result, there should be a gradual transition in the index as one goes 
from objects with an effective temperature of about 1800 K to cooler 
temperatures.
Therefore the $K1-K2$ plot should be a good indicator of the temperature 
of the object.

Values of K1 and K2 for various low-mass objects are shown in Table 2 
and in Figure~4.
\placetable{table2}
\placefigure{fig4}
We can see that cooler objects tend to be located toward the 
lower right of Figure 4.
Kelu-1 and D12-15  have strong Li absorption at 670.8 nm and 
are therefore considered to be confirmed brown dwarfs.
D02--11 and GD165B are considered to be likely brown dwarfs and
are located at $K1$ $\ge$ 0.11.
The status of D10--15, whether a brown dwarf or not is unclear.

Several trends can be noted in Table 2 and Figure 4.
First, from warmest to coolest the DENIS objects have been previously ranked 
as D10--15, D12--15, and D02--11 by Tinney et al. (1997).
This is consistent with the trend seen in the K index plot where these 
objects progress to larger K1 index and smaller K2 index with lower 
effective temperature (see Figure 4).
This is easily seen as the result of deeper H$_2$O and H$_2$ absorption with 
lower effective temperature.
Second, GL229B characterized by an extremely negative K2 index but with a K1 
index which is typical of much warmer objects.  
We therefore expect that the K2 index to be more important in distinguishing 
objects cooler than D02--11.

With more objects, it may be possible to establish a relationship between 
the effective temperature of the object and the position on the $K1-K2$ 
plot.
At the present time the effective temperatures for D10--15, D12--15, and 
D02--11 are very uncertain.
Detailed models, such as those presented by Tsuji et al. (1996), is needed 
to establish the effective temperatures of these objects.
Until this is accomplished, the effective temperatures of these objects 
should be treated with a great deal of caution.
In the case of D02--11, the effective temperature is estimated to
be about 1800 K but not as cool as 1500 K.
Note that atmospheric models of brown dwarfs have effective temperature
uncertainties of 100--200~K (Ruiz et al. 1997, Tsuji et al. 1996).
It is also important to keep in mind that dust formation is an important 
factor in these atmospheric models (Jones \& Tsuji 1997; Tsuji et al. 1996).
This adds further complication in seeking a straightforward interpretation 
using the K index.

\subsection{Photometry}

Our narrowband photometry provides the more precise information on the
continuum level of these objects compared to that presented by
Delfosse et al. (1997).  
Compared to broadband photometry,
the narrowband data is largely unaffected by the 
H$_2$O absorption because it is centered on the continuum and it is
insensitive to color transformation uncertainties that exist
in broadband photometric observations.

Figure 5 shows the photometry and a comparison to blackbody curves and 
to broadband photometry.
\placefigure{fig5}
Because of the strong H$_2$O absorption bands,
the broadband magnitudes are 0.1--0.4 mag lower
than the narrowband magnitudes.
Although a single color temperature can fit the 1--3.8 $\mu$m photometry,
there is little physical significance to the color temperature 
because these objects do not radiate like blackbodies. 
Models by Tsuji et al. (1996) show the extremely nonblackbody 
emission of low-temperature objects.
In addition one can see in Figure 5 that a single blackbody fails to match  
the 0.79 $\mu$m point.

\section{Conclusions}


From our 1.9--2.5 $\mu$m spectra and narrowband photometry, we 
conclude:

1. A simple two-parameter spectral slope analysis shows promise as a 
means of estimating the effective temperature of low-mass objects 
(see Fig. 4).
Brown dwarfs are likely to have $K1$ $\ge$ 0.25. 

2. D02--11 is the coolest of the objects reported by Delfosse et al. (1977),
and its spectral shape appears to be intermediate between that of 
D10--15 and GL229B.

3. We present accurate accurate narrowband photometry in the continuum
for the first time.  
This will provide better constraints on models of very low mass objects, 
especially in providing an estimate of the effective temperature.

\acknowledgments{
We thank M. Cushing for assistance in data reduction and T. Kerr for 
assistance at UKIRT and removal of the bad column data.  
We are also indebted to T. Tsuji for discussions about the spectra of 
low-mass objects.
}


\clearpage
 
\begin{deluxetable}{ccccc}
\footnotesize
\tablecaption{Narrowband  photometry  magnitudes  \label{table1} }
\tablewidth{4.0in}
\tablehead{   
\colhead{Object} & \colhead{[1.28]}   & \colhead{[1.68]}   & 
\colhead{[2.18]} & \colhead{[L$^\prime$]}    
} 
\startdata
D02--11 & 14.22 & 13.36 & 12.88  & 12.05 \nl
D10--15 & 13.88 & 13.08 &12.47 & 12.00  \nl
D12--15 & 14.01 & 13.16  &12.60  & 11.76 \nl
\enddata


\end{deluxetable}

\clearpage

\begin{deluxetable}{lccc}
\footnotesize
\tablecaption{K index for selected low-mass objects \label{table2} }
\tablewidth{5.5in}
\tablehead{   
\colhead{Object} & \colhead{K1}   & \colhead{K2}   & 
\colhead{T$_{eff}(K)$}
} 
\startdata
GL 406  & 0.0257 & --0.0244 & 2580\tablenotemark{a}\nl
LHS 2924 & 0.0880 & --0.0313 & 2080\tablenotemark{a}\nl
VB 10   & 0.100 & --0.00957 & 2330\tablenotemark{a}\nl
D10--15 & 0.226 & --0.0264 & 1800?\tablenotemark{b}\nl
BRI 0021--0214 & 0.225 & 0.0331 & 1980\tablenotemark{a}\nl
D12--15 & 0.299 & --0.0694 & 1600?\tablenotemark{c}\nl
Kelu 1  & 0.255 & --0.00662 & 1900\tablenotemark{d}\nl
GD 165B & 0.309 & --0.00426 & 1860\tablenotemark{e}\nl
GL 229B & 0.320 & --1.172 & 1000\tablenotemark{f} \nl
D02--11 & 0.384 & --0.153 & 1800?\tablenotemark{g}\nl

\enddata

\tablenotetext{a}{Tinney et al. (1993).  The value shown is the ``equivalent 
temperature'' from observations.}
\tablenotetext{b}{Delfosse et al. (1997). Conjecture.}
\tablenotetext{c}{Mart\'{\i}n et al. (1997). Conjecture.
    Object has Li absorption.}
\tablenotetext{d}{Ruiz et al. (1997).  
    Based on atmospheric model.  Object has Li absorption.}
\tablenotetext{e}{Jones et al. (1994).  Based on observations.}
\tablenotetext{f}{Tsuji et al. (1996).  Based on a atmospheric model.
    Object has very strong CH$_4$ absorption.}
\tablenotetext{g}{This work.  Based on atmospheric model by Tsuji (1998),
    but uncertain.}

\end{deluxetable}

%
%

\clearpage

\figcaption [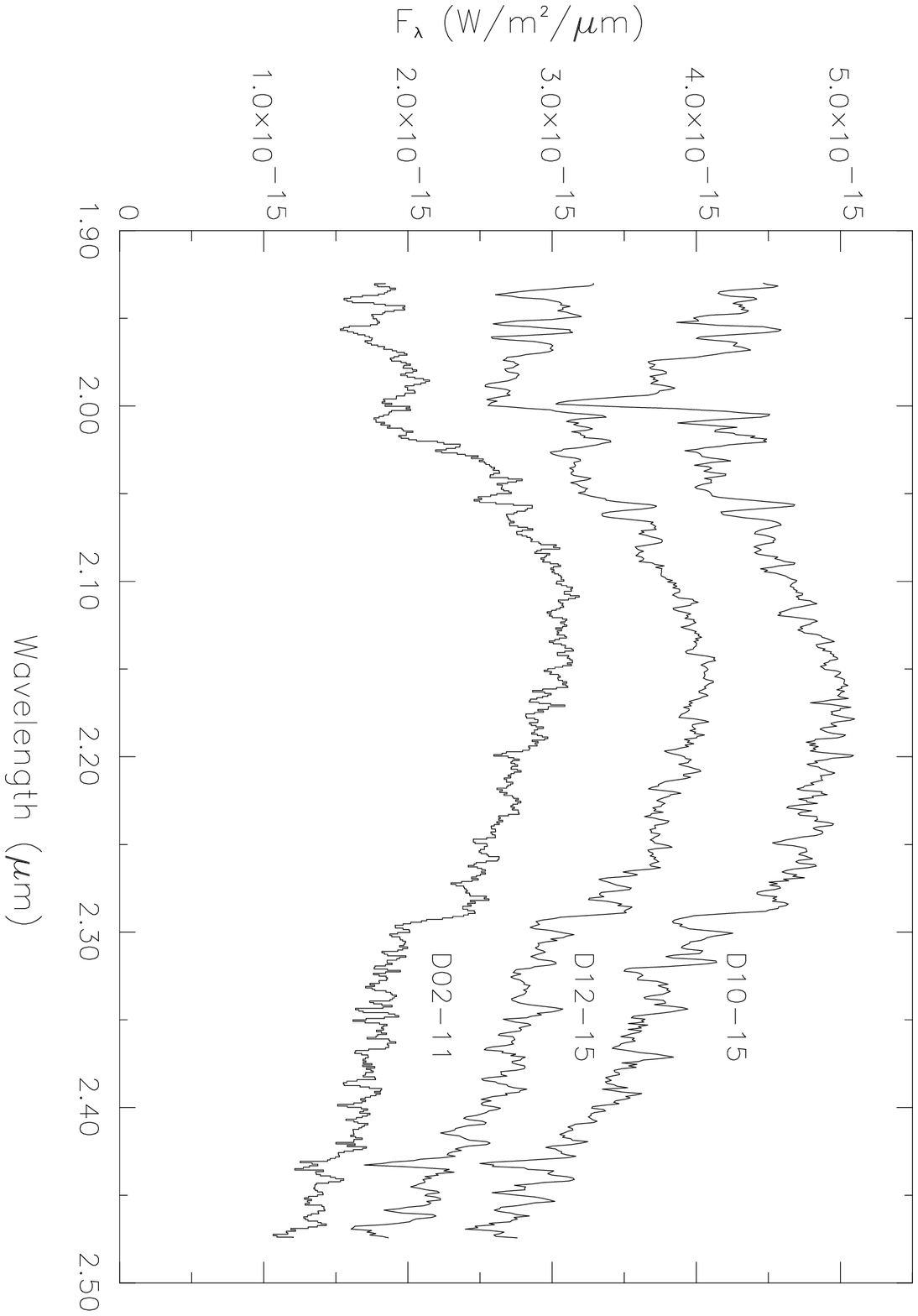] {Spectra of the DENIS objects 
in order of warmer to colder.
For clarity, the spectrum of D12--15 was multiplied by 2.0 and that of
D10--15 was multiplied by 4.0.   The possible Na I absorption feature
is marked. \label{fig1} } 

\figcaption [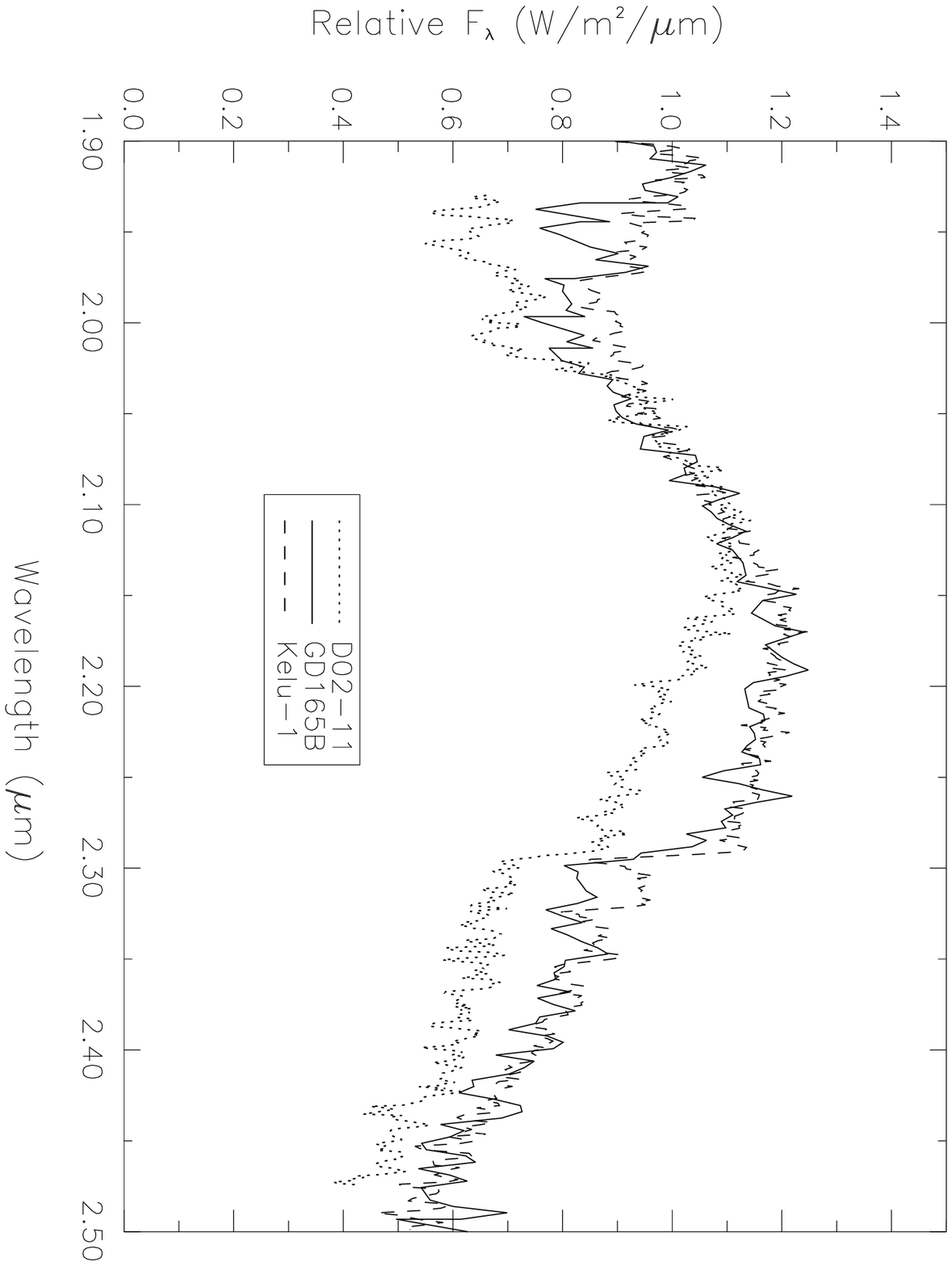] {Comparison of D02--11 to Kelu-1 
and GD165B.  Note the 
similarity of Kelu-1 (Ruiz et al. 1997) and GD165B (Jones et  al. 1994).  
Also, there is a pronounced dip in the spectrum of D02--11 at 
2.15--2.30 $\mu$m compared to the other two objects.  
The spectra were normalized at 2.07 $\mu$m.  \label{fig2} }

\figcaption [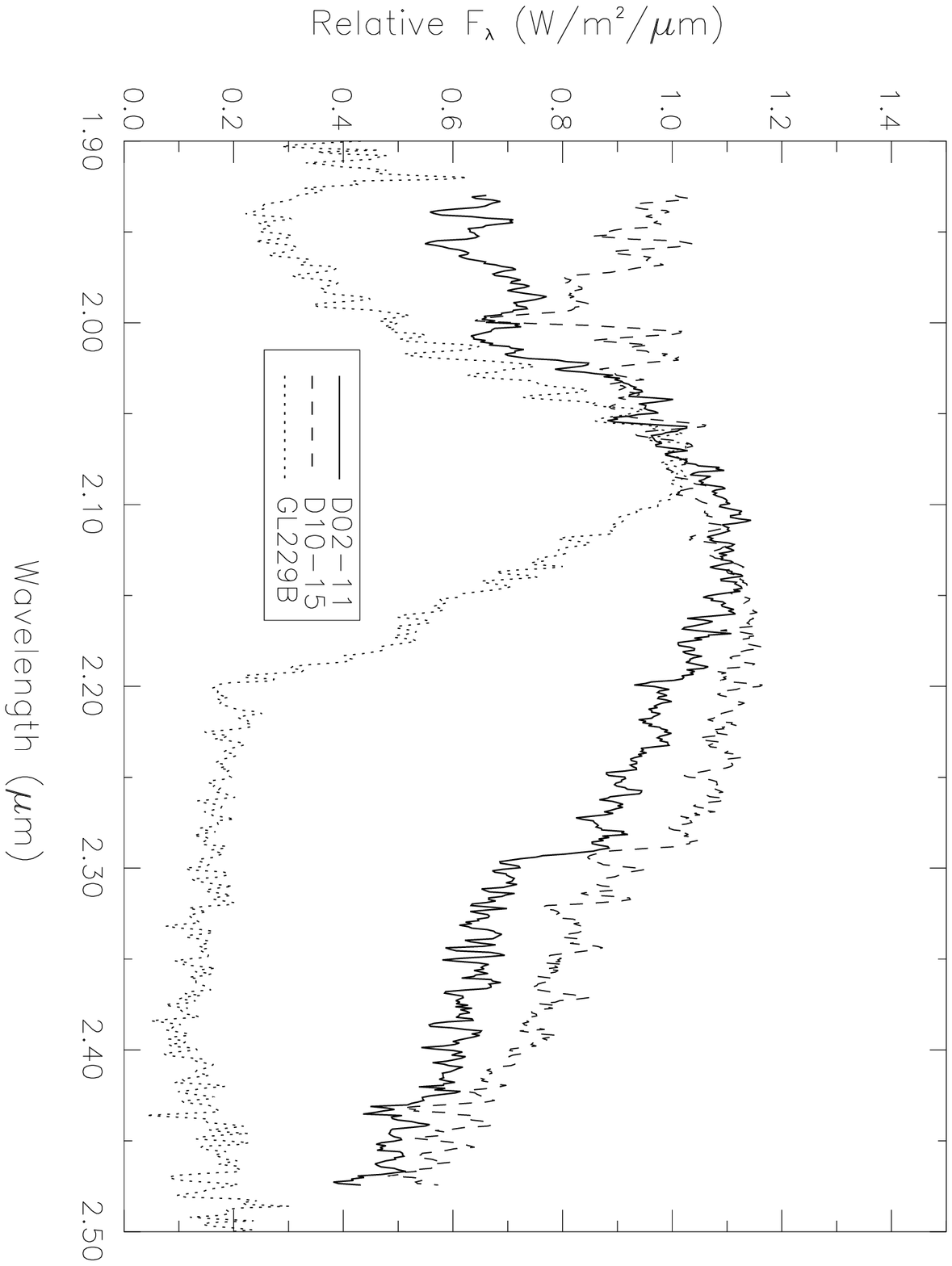] {Comparison of D02--11 to D10--15 and 
GL229B (Geballe et al. 1998).
The spectra were normalized at 2.07 $\mu$m.  \label{fig3} }

\figcaption [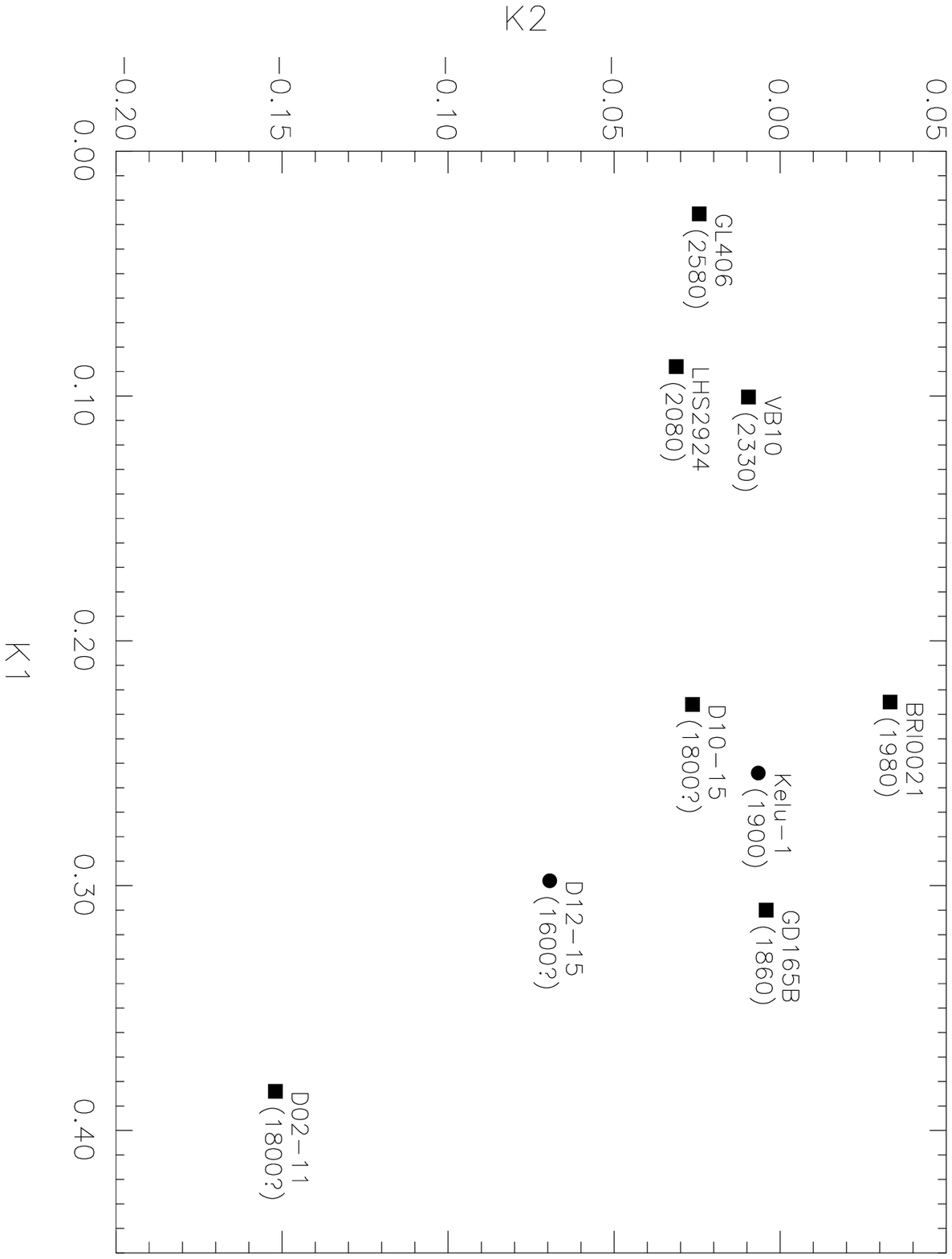] {Plot of the two $K$ indices.  
The estimated effective temperatures are shown in parentheses.  
Effective temperatures with ``?'' implies that these values are based on 
conjecture rather than observations or atmospheric models.
The solid circles indicate objects which have Li absorption.
\label{fig4} }

\figcaption [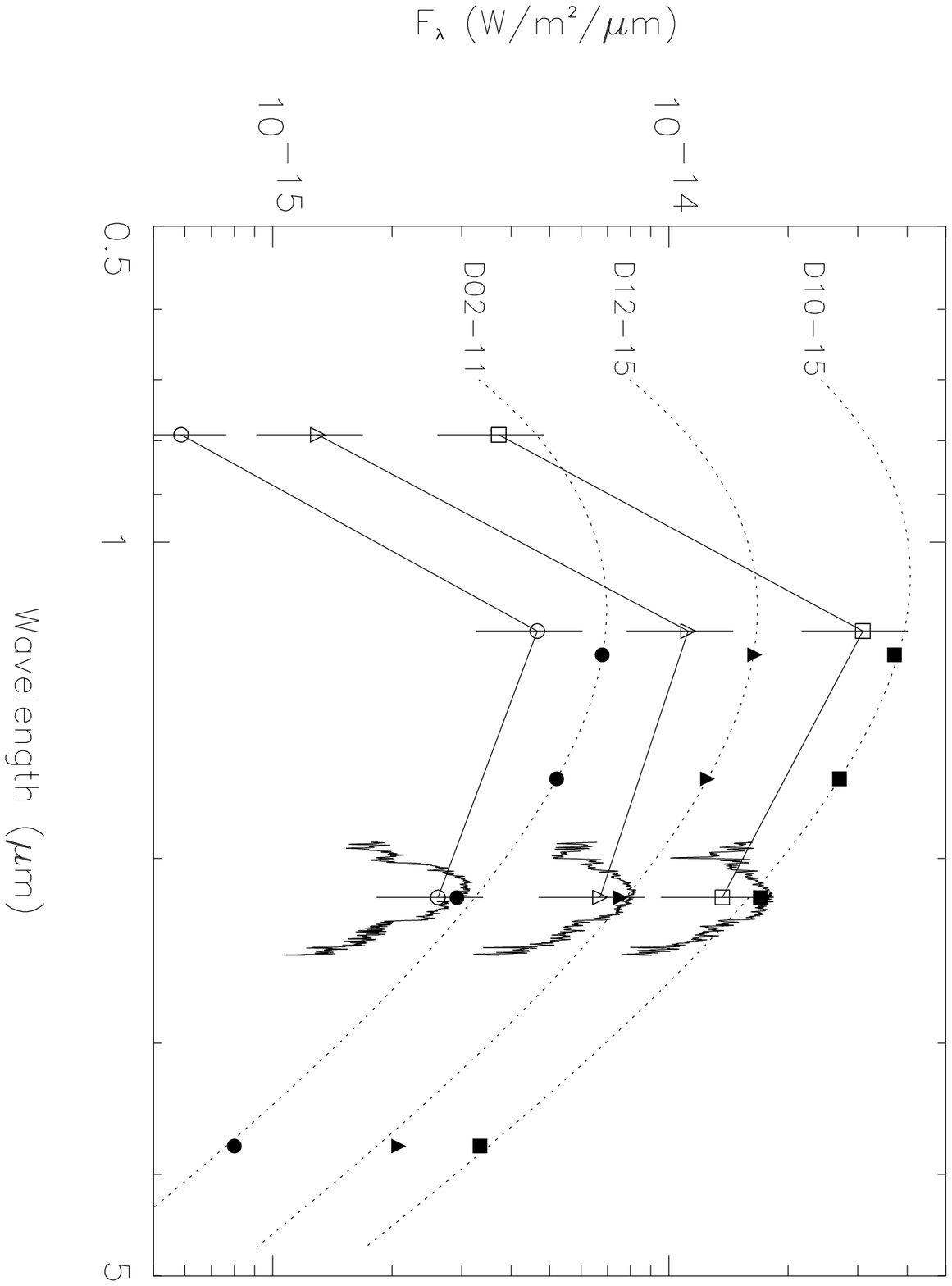] {Comparison of narrowband and 
broadband photometry.  
The solid symbols show our narrowband photometry.  
The open symbols show the broadband photometry reported  by 
Delfosse et al. (1997).
The solid line shows blackbody curves normalized to the 1.68 
$\mu$m narrowband photometric point.
The temperatures assumed were 2500 K, 2500 K, and 2700 K for D02--11, 
D12--15, and D10--15, respectively.  \label{fig5} }

\clearpage

\end{document}